\documentclass[nofootinbib,twocolumn,showpacs,preprintnumbers,pre,aps]{revtex4-1}

\usepackage[dvipdfmx]{graphicx}
\usepackage{bm}
\usepackage{amsmath}
\usepackage{amssymb}
\usepackage{color}

\begin{document}

\title{Irreversibility of stochastic state transitions in Langevin systems with odd elasticity}%

\author{Kento Yasuda}\email{yasudak@kurims.kyoto-u.ac.jp}
\affiliation{
Research Institute for Mathematical Sciences, Kyoto University, Kyoto 606-8502, Japan}

\date{\today}

\begin{abstract}
Active microscopic objects, e.g., an enzyme molecule, are modeled by the Langevin system with the odd elasticity, in which energy injection from the substrate to the enzyme is described by the antisymmetric part of the elastic matrix.
By applying the Onsager--Machlup integral and large deviation theory to the Langevin system with odd elasticity, we can calculate the cumulant generating function of the irreversibility of the state transition.
For an $N$-component system, we obtain a formal expression of the cumulant generating function and demonstrate that the oddness $\lambda$, which quantifies the antisymmetric part of the elastic matrix, leads to higher-order cumulants that do not appear in a passive elastic system.
To demonstrate the effect of the oddness under the concrete parameter, we analyze the simplest two-component system and obtain the optimal transition path and cumulant generating function. The cumulants obtained from expansion of the cumulant generating function increase monotonically with the oddness.
This implies that the oddness causes the uncertainty of stochastic state transitions.\end{abstract}

\maketitle

\section{Introduction}
\label{Intro}

In active matter, e.g., birds, bacteria convert chemical energy into mechanical work. To describe this energy injection by the material constants, odd elasticity, i.e., an antisymmetric part of an elastic tensor, is introduced~\cite{Scheibner20,Fruchart22}.
The odd elasticity breaks the Maxwell--Betti reciprocity, which holds on the elastic body conserving the mechanical energy. In other words, the odd elasticity quantifies the energy injection. 
Originally, odd elasticity was introduced for two-dimensional isotropic solids~\cite{Scheibner20} and was then extended to plates of moderate thickness~\cite{Fossati22}, a deformable membrane~\cite{AlIzzi23}, and linkage systems with active hinges~\cite{Ishimoto22,Brandenbourger22}.
Currently, odd elasticity is applied to several biological systems, e.g., starfish embryos~\cite{Tan22}, human sperm~\cite{Ishimoto23}, and muscles~\cite{Shankar22}, to capture their activity.
In addition, odd elasticity is implemented programmatically in artificial robots~\cite{Brandenbourger22}.

Odd elasticity has the potential to describe the structural dynamics of active enzymes, which change their structure during catalytic reactions from the substrate to products or nonreactive binding with inhibiting molecules~\cite{Togashi10,Toyabe15,Dey16,Brown20,Mugnai20,Yasuda21a,HK22,Toyabe10}.
In the case of catalytic reactions, energy is injected into the enzymes from the substrate.
We suggest that odd elasticity can model this energy injection and propose the model given by a Langevin system that includes odd elasticity~\cite{YIKLSHK22,YKLHSK22,KYILSHK23}.

Irreversibility is the ratio between the probability of a forward path and a time-reversed path, and it has been quantified in various non-equilibrium systems.
In stochastic thermodynamics, the irreversibility becomes thermodynamic entropy production and characterizes time-irreversibility of the whole system~\cite{Seifert05,Seifert12}.
Concrete calculations of the entropy production for specific Langevin systems have been performed.
The steady-state entropy production of the charged Brownian particle in a magnetic field has been calculated in several studies~\cite{Aquino10,Jayannavar07,Saha08,Aquino09}.
Weiss proposed a general theory of the linear Langevin system and calculated the cumulants of irreversibility in the steady-state~\cite{Weiss07}.

Some active matter models are represented by only macroscopic variables, and the microscopic degrees of freedom are coarse-grained.
Hence, the entropy produced by the microscopic degrees of freedom is lost, and the correspondence between the irreversibility and the thermodynamic entropy production is broken~\cite {Fodor22,Byrne22}.
Even if the irreversibility does not represent the thermodynamic entropy production, it is still an important quantity to discuss the phenomenological behaviors of active systems.
A typical problem showing the utility of the irreversibility is the locomotion of an active element in overdamped systems.
It is known that the Scallop theorem states reciprocal (reversible) deformations can not be used for locomotion in the overdamped systems~\cite{Purcell77,Ishimoto12}.
As a consequence of the Scallop theorem, locomotion velocity is proportional to the irreversibility of deformations~\cite{Shapere89}.
A theoretical framework based on the Scallop theorem is applied wide range of active systems such as active transport of an enzyme molecule~\cite{Yasuda21a,Golestanian15,Sakaue10,Vishen24}, cell swimming in viscous fluids~\cite{Lauga09}, cell crawling~\cite{Tarama18,Leoni17}, and macroscopic robots moving on sand~\cite{Hatton13}.

Here, we consider the stochastic state transition between the initial and final states $\mathbf x^\mathrm i\to\mathbf x^\mathrm f$ with duration time $t_\mathrm f$. 
For example, this transition models the conformation change of an enzyme induced by combining the substrate molecule.
To extract the statistical properties of these state transitions, we can use the path integral formalism of a stochastic system called the Onsager--Machlup theory~\cite{Onsager53,Tomita74b,RiskenBook,ZuckermanBook,Doi19,Wang21,FK82,TW80,Taniguchi07,Taniguchi08}. In this theory, the probability of observing each trajectory is given with a quantity referred to as the Onsager--Machlup integral and is used to extract the most probable path~\cite{Durr,Wissel79,Faccioli06,Adib08,Wang10,Gladrow,YKLHSK22}. 
Note that the most probable path of the active particles has attracted significant interest recently~\cite{Teeffelen08,Woillez19,Majumdar20,Gu20,Yasuda22b}.
This optimization problem is also used to calculate the cumulant generating functions in the framework of the large deviation theory~\cite{Touchette09,Krapivsky14,Mallick22,Falasco23}.

In this paper, we calculate the cumulant generating functions of the irreversibility of the state transition $\mathbf x^\mathrm i\to\mathbf x^\mathrm f$ in the Langevin system with the odd elasticity.
To perform this calculation, we apply the Onsager--Machlup theory to the $N$-component Langevin system with the odd elasticity.
In addition, by solving the optimization program of the Onsager--Machlup integral under the large deviation theory and Varadhan's theorem~\cite{Varadhan66,Touchette09}, regarded as the saddle point approximation, we obtain a formal expression of the cumulant generating functions of the irreversibility. 
As a result, we found that the odd elasticity leads to the variance of the irreversibility, which disappear in the case of a purely passive system.
In addition, we perform a specific analysis on a two-component system and demonstrate the concrete expression of the optimal trajectories and cumulant generating functions.

Some previous reports calculated the cumulant generating functions without using the saddle point approximation in a linear Langevin system with an antisymmetric drift force~\cite{Kwon11,Buisson23,Chernyak06,Turitsyn07,Noh13,Noh14}, which is mathematically same as the Langevin system with the odd elasticity.
Unlike these previous papers that focused on steady-state or time series under an unfixed initial or final state, this paper focuses on the state transition with fixed initial and final states.
Calculations for the state transition are more complicated than the steady-state, because the state transition is a two-time boundary problem with conditions at two different time points. 
Hence, as an initial attempt, we employ the saddle point approximation to calculate the cumulant generating functions for the state transition.

The remainder of this paper is organized as follows. In Section~\ref{Model}, we present the formalism of the Langevin system with the odd elasticity and the main formal results of the cumulant generating function of the irreversibility.
Section~\ref{TCS} calculates the cumulant generating function in the simplest two-component system to demonstrate the drastic influence of oddness on the entropy change.
In Section~\ref{Enzyme}, we show the relation between an enzyme and our theory.
Finally, Section~\ref{Dis} provides a summary and relevant discussions.

\section{Langevin system with odd elasticity}
\label{Model}
\begin{figure}[t]
\begin{center}
\includegraphics[scale=0.35]{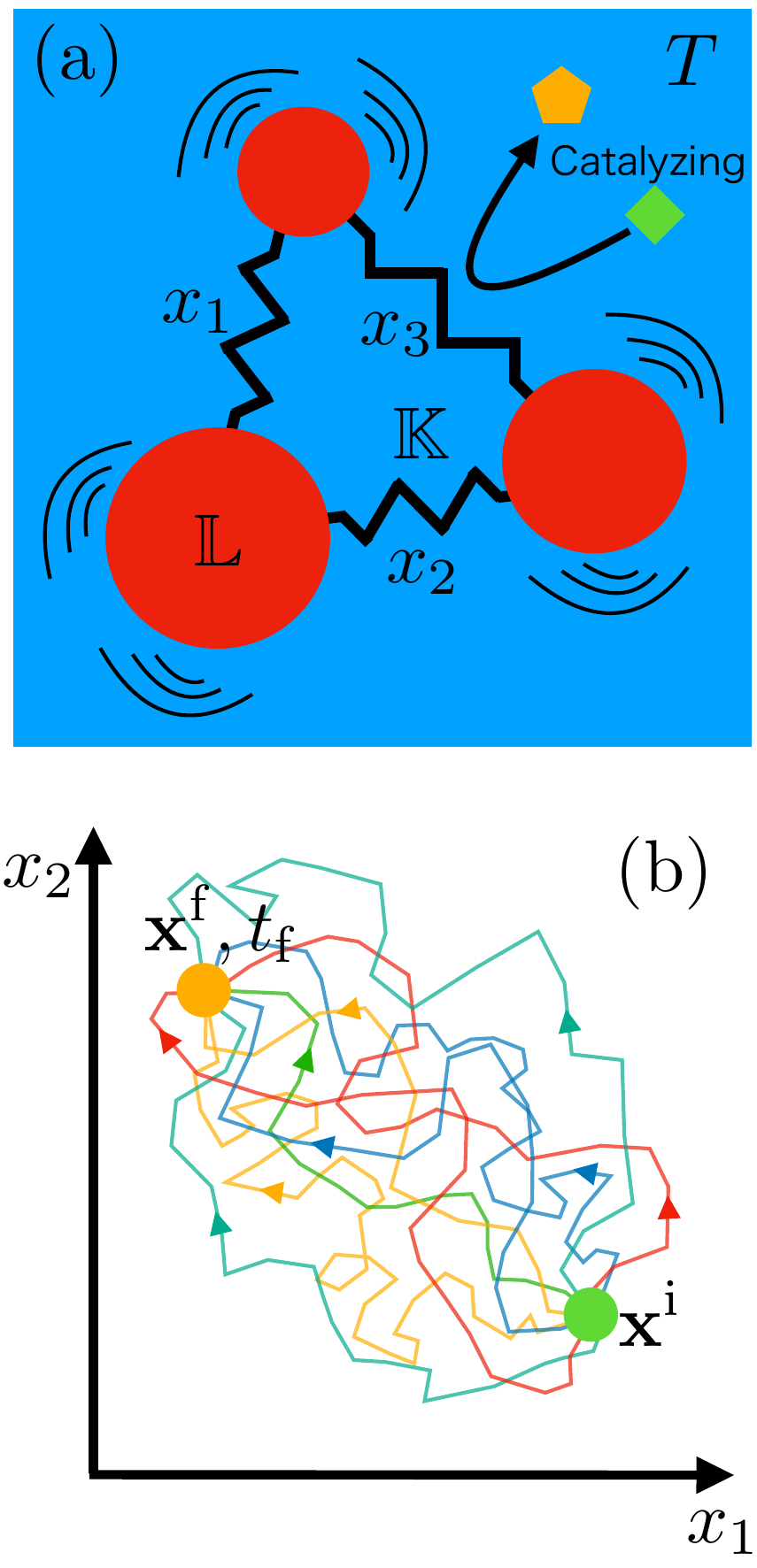}
\end{center}
\caption{
(a) Sketch of an object described by the Langevin system with the odd elasticity Eq.~(\ref{OddLangevinSystem})~\cite{YKLHSK22,YIKLSHK22}.
The object consists of domains (red circles) connected by springs embedded in the dissipative environment, e.g., a viscous fluid, with temperature $T$. The structure of the object is represented by the set of the spring lengths $\mathbf x=(x_1,x_2,\cdots,x_N)^\mathrm T$.
The mobility of the domains is characterized by the mobility matrix $\mathbb L$, and the stiffness of springs is characterized by the elastic matrix $\mathbb K$.
The object catalyzes a chemical reaction from a substrate molecule (green square) to a product molecule (orange pentagon).
The energy the object receives from the substrate molecule through the reaction is modeled by the odd elasticity $\lambda \mathbb A$.
The domains experience fluctuation caused by the thermal noise of the dissipative environment.
(b) Stochastic transitions from the initial state $\mathbf x^\mathrm i$ (green circle) to final state $\mathbf x^\mathrm f$ (orange circle) with duration time $t_\mathrm f$ in a two-dimensional state space. Each trial represented by the trajectories connecting $\mathbf x^\mathrm i$ and $\mathbf x^\mathrm f$ is random and differs from the other trials.
}
\label{Fig:model}
\end{figure}

In our recent study, we introduced a linear Langevin system with odd elasticity~\cite{YKLHSK22,YIKLSHK22} to capture an aspect of a catalytic enzyme, which is transferred with energy by chemical reactions (Fig.~\ref{Fig:model} (a)).
We showed that the odd elasticity partially describes the nonreciprocal conformational dynamics of the enzyme~\cite{KYILSHK23}.
In the following, we describe the formalism of the $N$-component  Langevin system and calculate the cumulant generating functions of the irreversibility.

\subsection{$N$-component Langevin system with the odd elasticity}

Here, we consider the $N$-element vector $\mathbf x$ whose components are $x_i ~(i=1,2\cdots N)$, which obey following Langevin system~\cite{YKLHSK22,YIKLSHK22,KYILSHK23}:
\begin{align}
\dot{\mathbf x}=-\ell k\mathbb L \mathbb K \mathbf x+\mathbb F\mathbf y(t),
\label{OddLangevinSystem}
\end{align}
where the dot represents the time derivative, $\dot{\mathbf x}=d\mathbf x/dt$. Here, $\ell$ is a scalar with a mobility dimension, and $\mathbb L$ is an $N\times N$ nondimensional mobility matrix. According to Onsager's reciprocal theorem and the second law of thermodynamics, $\mathbb L$ is a symmetric-positive definite matrix~\cite{KuboBook,DoiBook}. In addition, $k$ is a scalar having the dimension of a spring constant, $\mathbb K$ is an $N\times N$ nondimensional elastic matrix that can be constructed by symmetric and antisymmetric parts~\cite{YKLHSK22,YIKLSHK22,KYILSHK23}, i.e., $\mathbb K=\mathbb S+\lambda \mathbb A$, where $\mathbb S^\mathrm T=\mathbb S$, $\mathbb A^\mathrm T=-\mathbb A$, and the superscript $\mathrm T$ represents transpose.
The symmetric and antisymmetric parts are regarded as conservative and nonconservative forces, respectively.
$\lambda$ is a scalar representing the magnitude of the antisymmetric part, which can be referred to as ``oddness." For the stability of this system, $\mathbb S$ is assumed to be positive definite. The second term on the right-hand side of Eq.~(\ref{OddLangevinSystem}) represents thermal fluctuations characterized by Gaussian white noise $\mathbf y$ satisfying $\langle \mathbf y(t)\rangle=0$ and $\langle \mathbf y(t)\mathbf y(0)\rangle=\mathbb I_N\delta(t)$, where $\langle \cdot\rangle$ indicates the statistical average, $\mathbb I_N$ is an $N\times N$ identity matrix, and $\delta (t)$ is Dirac's delta function. Note that the $N\times N$ amplitude matrix $\mathbb F$ obeys the fluctuation-dissipation relation~\cite{KuboBook,DoiBook} $\mathbb F^\mathrm T\mathbb F=2k_\mathrm BT\ell\mathbb L$, where $k_\mathrm B$ is the Boltzmann constant, $T$ is the temperature of the thermal bath. The relaxation rate of this system is given as $\gamma=\ell k$.

The path probability of the trajectory obeying Eq.~(\ref{OddLangevinSystem}), $\mathbf x(t)$, which begins at $\mathbf x^\mathrm i$ until $t=t_\mathrm f$, is given as follows~\cite{Onsager53,RiskenBook}:
\begin{align}
P[\mathbf x(t)|\mathbf x^\mathrm i]=\mathcal N \exp[-O/(2k_\mathrm BT)].
\label{PathProb}
\end{align}
Here, $O$ is the Onsager--Machlup integral given by~\cite{YKLHSK22,Tomita74b}:
\begin{align}
O=\frac{k}{2}\int_0^{t_\mathrm f} dt\gamma\, [\mathbf v\gamma^{-1}+\mathbb L\mathbb K\mathbf x]^\mathrm T\mathbb L^{-1}[\mathbf v\gamma^{-1}+\mathbb L\mathbb K\mathbf x],
\label{OMIntegral}
\end{align}
where $\mathbf v=\dot{\mathbf x}$, and $\mathcal N$ is a normalizing constant.

\subsection{Irreversibility of state transitions and its cumulant generating function}
Here, we consider the stochastic transitions from the initial state $\mathbf x^\mathrm i$ to the final state $\mathbf x^\mathrm f$ with duration time $t_\mathrm f$. As shown in Fig.~\ref{Fig:model}(b), the trajectory connecting $\mathbf x^\mathrm i$ and $\mathbf x^\mathrm f$ is independent in each trial because overall the system is stochastic.
To characterize the transition $\mathbf x^\mathrm i\to\mathbf x^\mathrm f$, we consider the irreversibility.
Generally, the irreversibility depends on $\mathbf x^\mathrm i$ and $\mathbf x^\mathrm f$, as well as the trajectory $\mathbf x(t)$ during the transition.
We define the irreversibility as $\Sigma:= k_\mathrm B\ln(P[\mathbf x(t)|\mathbf x^\mathrm i]/P[\mathbf x^{\mathrm {rev}}(t)|\mathbf x^\mathrm f])$~\cite{Seifert12}, where we introduce the reversed trajectory $\mathbf x^{\mathrm {rev}}(t)=\mathbf x(t_\mathrm f-t)$. 
Using Eq.~(\ref{OMIntegral}), the irreversibility $\Sigma$ of our system is expressed as follows:
\begin{align}
T\Sigma =-\Delta U-k\lambda \int_0^{t_\mathrm f} dt \mathbf v^\mathrm T\mathbb A \mathbf x.
\label{Entropy}
\end{align}
Here, $\Delta U$ is the difference in the elastic energy between the initial and final states:
\begin{align}
\Delta U:=\frac{k}{2}[(\mathbf x^\mathrm f)^\mathrm T\mathbb S\mathbf x^\mathrm f-(\mathbf x^\mathrm i)^\mathrm T\mathbb S\mathbf x^\mathrm i]
\label{ElasticEnergy},
\end{align}
including only the symmetric part of the elastic matrix $\mathbb S$.
In the passive system $\lambda=0$, the irreversibility corresponds to  $\Delta U$ without an error. 
Hence, before performing a detailed calculation, it is expected that the oddness $\lambda$ causes fluctuation in $\Sigma$, which is quantified by the second cumulant $\langle\Sigma^2\rangle_\mathrm c$.

To characterize the stochastic property of the irreversibility, we introduce the cumulant generating function $\Phi(\xi):=\ln\langle e^{\xi\Sigma/k_\mathrm B}\rangle_{\mathrm i\to \mathrm f}$~\cite{KardarBook}, where $\langle \circ \rangle_{\mathrm i\to \mathrm f}$ indicates the statistical average under the condition of the initial and final state:

\begin{align}
\langle \circ \rangle_{\mathrm i\to \mathrm f}:=\int_{\mathbf x^\mathrm i}^{\mathbf x^\mathrm f} \mathcal D\mathbf x\,\circ P[\mathbf x(t)|\mathbf x^\mathrm i,\mathbf x^\mathrm f].
\end{align}
Here, $P[\mathbf x(t)|\mathbf x^\mathrm i,\mathbf x^\mathrm f]$ is a path probability under the condition that $\mathbf x(0)=\mathbf x^\mathrm i$ and $\mathbf x(t_\mathrm f)=\mathbf x^\mathrm f$, and can be given by Bayes' theorem $P[\mathbf x(t)|\mathbf x^\mathrm i,\mathbf x^\mathrm f]\mathcal P(\mathbf x^\mathrm f|\mathbf x^\mathrm i)=P[\mathbf x(t)|\mathbf x^\mathrm i]$, where $\mathcal P(\mathbf x^\mathrm f|\mathbf x^\mathrm i)$ is a probability of $\mathbf x^\mathrm f$ at $t_\mathrm f$ under the condition $\mathbf x(0)=\mathbf x^\mathrm i$.
In addition, $\int_{\mathbf x^\mathrm i}^{\mathbf x^\mathrm f} \mathcal D\mathbf x$ indicates the path integral over all trajectories $\mathbf x(t)$ satisfying $\mathbf x(0)=\mathbf x^\mathrm i$ and $\mathbf x(t_\mathrm f)=\mathbf x^\mathrm f$.
According to Eq.~(\ref{PathProb}), the cumulant generating function of the irreversibility is rewritten as follows:
\begin{align}
&\Phi(\xi)=C+\ln\int_{\mathbf x^\mathrm i}^{\mathbf x^\mathrm f} \mathcal D\mathbf x\exp[-\hat O/(2k_\mathrm BT)],
\label{CGF}
\end{align}
where $C$ is a $\xi$-independent constant determined by a normalization condition $\Phi(0)=0$, and $\hat O$ is the modified Onsager--Machlup integral:
\begin{align}
&\hat O:=O-2\xi T\Sigma \nonumber\\
&=(1+2\xi)\Delta U
+\frac{k}{2}\int_0^{t_\mathrm f} dt\gamma\,\begin{pmatrix}
    \mathbf x\\\mathbf v\gamma^{-1}
    \end{pmatrix}^\mathrm T(\mathbb R_0+\xi\lambda \mathbb R_1)\begin{pmatrix}
    \mathbf x\\\mathbf v\gamma^{-1}
    \end{pmatrix}.
    \label{ModOMIntegral}
    \end{align}
Here, the following $2N\times 2N$ matrixes are introduced:
\begin{align}
\mathbb R_0:=\begin{pmatrix}
    \mathbb K^\mathrm T\mathbb L \mathbb K&-\lambda\mathbb A\\\lambda\mathbb A&\mathbb L^{-1}
    \end{pmatrix},~~~
    \mathbb R_1:=\begin{pmatrix}
    \mathbb O_N&-2\mathbb A\\2\mathbb A&\mathbb O_N
    \end{pmatrix},
\end{align}
where $\mathbb O_N$ is an $N\times N$ zero matrix.
Note that all cumulants $\langle\Sigma^n\rangle_\mathrm c$ are obtained from $\Phi(\xi)$ as~\cite{KardarBook}:
\begin{align}
\langle\Sigma^n\rangle_\mathrm c=k_\mathrm B^n\left.\frac{d^n}{d\xi^n}\Phi(\xi)\right|_{\xi=0}.
\label{CumulantsQGF}
\end{align}

Under the large deviation theory and Varadhan's theorem~\cite{Varadhan66,Touchette09}, the cumulant generating function is approximated as follows:
\begin{align}
&\Phi(\xi)-C\sim -\inf_{\mathbf x(t)|\mathbf x^\mathrm i,\mathbf x^\mathrm f}[\hat O]/(2k_\mathrm BT).
\label{VaradhanT}
\end{align}
under the condition that $\hat O/(k_\mathrm BT)\gg1$.
This means that we can calculate the cumulant generating function by considering the optimization problem of $\hat O$ with respect to the trajectory $\mathbf x(t)$ under the condition that $\mathbf x(0)=\mathbf x^\mathrm i$ and $\mathbf x(t_\mathrm f)=\mathbf x^\mathrm f$.
That indicates Eq.~(\ref{VaradhanT}) can be regarded as a saddle point approximation of the path integral.

In this study, we found that the cumulant generating function of irreversibility of state transition $\mathbf x^\mathrm i\to\mathbf x^\mathrm f$ with duration time $t_\mathrm f$ is given in the following form:
\begin{align}
&\Phi(\xi)=-\frac{\xi \Delta U}{k_\mathrm BT}-\frac{k}{2k_\mathrm BT}\sum_{n=1}\frac{\xi^n\lambda^n}{2}
\begin{pmatrix}
    \mathbf x^\mathrm i\\\mathbf x^\mathrm f
    \end{pmatrix}^\mathrm T 
\mathbb E_n(\lambda,t_\mathrm f)
    \begin{pmatrix}
    \mathbf x^\mathrm i\\\mathbf x^\mathrm f
    \end{pmatrix},
    \label{ResultCGF}
\end{align}
where $\mathbb E_n$ are $2N\times 2N$ matrixes that characterize each cumulants.
Note that the formal expression of $\mathbb E_n$ is given in Eq.~(\ref{GeneralResultE}).
From this expression, we find that the oddness $\lambda$ contributes to the first cumulant and higher cumulants, which implies that the oddness leads to an uncertainty of the irreversibility.
In contrast, if the elastic matrix is symmetric ($\lambda=0$), we can obtain the cumulant generating function from Eq.~(\ref{ResultCGF}) as follows:
\begin{align}
\Phi(\xi)=-\xi\frac{\Delta U}{k_\mathrm BT}.
\label{Q0}
\end{align}
Thus, the first cumulant and higher cumulants are given as:
\begin{align}
&\langle \Sigma\rangle _\mathrm c=-\frac{\Delta U}{T},\\
&\langle \Sigma^2\rangle _\mathrm c=\langle \Sigma^3\rangle _\mathrm c=\cdots=0.
\end{align}
This means that the irreversibility corresponds to elastic energy change between the initial and final state without error.

\subsection{Extremum equation and its solutions}
In the following, we present the derivation of Eq.~(\ref{ResultCGF}) and a formal expression for the matrix $\mathbb E_n$.
Recalling Eq.~(\ref{VaradhanT}), we consider the variational problem of $\hat O$, i.e., $\delta \hat O=0$~\cite{Touchette09,Durr,YKLHSK22} to obtain $\inf_\mathbf x[\hat O]$. This requirement and Eq.~(\ref{ModOMIntegral}) lead to the following extremum equation:
\begin{align}
\dot{\mathbf v}\gamma^{-2}+2(1+2\xi)\lambda \mathbb L\mathbb A\mathbf v\gamma^{-1}-\mathbb L \mathbb K^\mathrm T\mathbb L \mathbb K\mathbf x=0.
    \label{pExtremumEq}
\end{align}
This equation can be rewritten as follows:
\begin{align}
\begin{pmatrix}
    \dot {\mathbf x}\gamma^{-1}\\\dot{\mathbf v}\gamma^{-2}
    \end{pmatrix}
=\mathbb{G}\begin{pmatrix}
    \mathbf x\\\mathbf v\gamma^{-1} 
    \end{pmatrix},
    \label{ExtremumEq}
\end{align}
where $\mathbb{G}$ is a $2N\times 2N$ matrix given as:
\begin{align}
\mathbb{G}:=\begin{pmatrix}
    \mathbb O_N&\mathbb I_N\\
    \mathbb L \mathbb K^\mathrm T\mathbb L \mathbb K&-2(1+2\xi)\lambda\mathbb L \mathbb A
    \end{pmatrix}.
\end{align}
To obtain the cumulants in order, we expand the variables as $\mathbf x=\sum_n\xi^n\mathbf x_n, \mathbf v=\sum_n\xi^n\mathbf v_n$ and obtain the equation for each component from Eq.~(\ref{ExtremumEq}) as follows:
\begin{align}
\begin{pmatrix}
    \dot {\mathbf x}_n\gamma^{-1}\\\dot{\mathbf v}_n\gamma^{-2}
    \end{pmatrix}
=\mathbb{G}_0\begin{pmatrix}
    \mathbf x_n\\\mathbf v_n\gamma^{-1}
    \end{pmatrix}+\lambda\mathbb{G}_1\begin{pmatrix}
    \mathbf x_{n-1}\\\mathbf v_{n-1}\gamma^{-1}
    \end{pmatrix},
    \label{ExpandedExtremumEq}
\end{align}
where we use the following $2N\times 2N$ matrixes:
\begin{align}
\mathbb{G}_0:=\begin{pmatrix}
    \mathbb O_N&\mathbb I_N\\
    \mathbb L \mathbb K^\mathrm T\mathbb L \mathbb K&-2\lambda\mathbb L \mathbb A
    \end{pmatrix},~
    \mathbb{G}_1:=\begin{pmatrix}
    \mathbb O_N&\mathbb O_N\\
    \mathbb O_N&-4\mathbb L \mathbb A
    \end{pmatrix}.
\end{align}

For $n=0$, we obtain the following formal solution to Eq.~(\ref{ExpandedExtremumEq}) with a matrix exponential:
\begin{align}
\begin{pmatrix}
    \mathbf x_0\\\mathbf v_0\gamma^{-1}
    \end{pmatrix}
    =e^{\mathbb G_0 \tau}
    \begin{pmatrix}
    \mathbf c_0\\\mathbf d_0
    \end{pmatrix},
\end{align}
where $\tau=t\gamma$ is a dimensionless time, and $\mathbf c_0$ and $\mathbf d_0$ are constants determined with the boundary conditions, i.e., $\mathbf x_0(0)=\mathbf x^\mathrm i$ and $\mathbf x_0(t_\mathrm f)=\mathbf x^\mathrm f$.
These boundary conditions can be rewritten in the following form:
\begin{align}
\begin{pmatrix}
    \mathbf x^\mathrm i\\0\\\mathbf x^\mathrm f\\0
    \end{pmatrix}
    =\mathbb B
        \begin{pmatrix}
    \mathbf c_0\\\mathbf d_0\\\mathbf v^\mathrm i\gamma^{-1}\\\mathbf v^\mathrm f\gamma^{-1}
    \end{pmatrix},
    \label{BC0}
\end{align}
where $\mathbf v^\mathrm i=\mathbf v_0(0)$ and $\mathbf v^\mathrm f=\mathbf v_0(t_\mathrm f)$ are also constants determined by the above boundary conditions, and $\mathbb B$ is a $4N\times 4N$ matrix given as follows:
\begin{align}
\mathbb B:=
    \begin{pmatrix}
     \mathbb I_{2N}&\mathbb J\\
    e^{\mathbb G_0 \tau_\mathrm f}&\mathbb J'
    \end{pmatrix},~
\mathbb J:
    =
    \begin{pmatrix}
     \mathbb O_N&\mathbb O_N\\
     -\mathbb I_N&\mathbb O_N
         \end{pmatrix},~
        \mathbb J':
    =
    \begin{pmatrix}
     \mathbb O_N&\mathbb O_N\\
     \mathbb O_N&-\mathbb I_N
         \end{pmatrix}.
   \end{align}
   By solving Eq.~(\ref{BC0}) for $\mathbf c_0$, $\mathbf d_0$, $\mathbf v^\mathrm i$, and $\mathbf v^\mathrm f$, we obtain the following:
\begin{align}
\begin{pmatrix}
    \mathbf c_0\\\mathbf d_0\\\mathbf v^\mathrm i\gamma^{-1}\\\mathbf v^\mathrm f\gamma^{-1}
    \end{pmatrix}=
    \mathbb B^{-1}
\begin{pmatrix}
    \mathbf x^\mathrm i\\0\\\mathbf x^\mathrm f\\0
    \end{pmatrix},
    ~~~
    \begin{pmatrix}
    \mathbf c_0\\\mathbf d_0
    \end{pmatrix}=
    \hat {\mathbb B}^{-1}
\begin{pmatrix}
    \mathbf x^\mathrm i\\\mathbf x^\mathrm f
    \end{pmatrix}.
    \end{align}
    Here, $\hat {\mathbb B}^{-1}$ is a $2N\times 2N$ matrix comprising $2N\times N$ matrixes $\mathbb C_n$, which are the block matrixes of $\mathbb B^{-1}$:
\begin{align}    
    \mathbb B^{-1}=\begin{pmatrix}
     \mathbb C_1&\mathbb C_2&\mathbb C_3&\mathbb C_4\\
     \mathbb C_5&\mathbb C_6&\mathbb C_7&\mathbb C_8\\
    \end{pmatrix}.
\end{align}
Using $\mathbb C_n$, we have 
$\hat{\mathbb B}^{-1}:=\begin{pmatrix}
     \mathbb C_1&\mathbb C_3    \end{pmatrix}$, and we use 
     $\tilde{\mathbb B}^{-1}:=\begin{pmatrix}
     \mathbb C_3&\mathbb C_4    \end{pmatrix}$ later.
Notably, $\mathbb B$ is a invertible except in the case  where $\tau_\mathrm f=0$. This is because $\det \mathbb B=\det [(e^{\mathbb G_0\tau_\mathrm f})_{12}]\ne0$, where $(e^{\mathbb G_0\tau_\mathrm f})_{12}$ is a upper-right submatrix of $e^{\mathbb G_0\tau_\mathrm f}$.

For arbitrary $n$, we have the following formal solution to Eq.~(\ref{ExpandedExtremumEq}):
\begin{align}
\begin{pmatrix}
    \mathbf x_n\\\mathbf v_n\gamma^{-1}
    \end{pmatrix}
    &=\sum_{m=0}^n\left[\lambda^me^{\mathbb G_0 \tau}\mathbb M_m(\tau)
    \begin{pmatrix}
    \mathbf c_{n-m}\\\mathbf d_{n-m}
    \end{pmatrix}\right],
    \label{solution}
\end{align}
where $\mathbf c_n$ and $\mathbf d_n$ are constants determined with the boundary conditions, i.e., $\mathbf x_n(0)=\mathbf x_n(t_\mathrm f)=0$. 
We also used the following functions:
\begin{align}
   & \mathbb M_m(\tau):=\int_0^{\tau} d\tau_1 \mathbb H(\tau_1)\cdots\int_0^{\tau_{m-1}} d\tau_m \mathbb H(\tau_m),\\
    &\mathbb H(\tau):=e^{-\mathbb G_0 \tau}
    \mathbb{G}_1e^{\mathbb G_0 \tau}.
\end{align}

Using the boundary conditions, we obtain the following relation (see also App.\ref{AppA}):
\begin{align}
    &\begin{pmatrix}
    \mathbf c_n\\\mathbf d_n
    \end{pmatrix}=\lambda^n\mathbb N_n
    \begin{pmatrix}
    \mathbf c_0\\\mathbf d_0
    \end{pmatrix},
    \label{SolutionBC}
\end{align}
where $\mathbb N_n$ is a matrix determined by the following recurrence relation:
\begin{align}
\mathbb N_n=\sum_{m=1}^n-\tilde{\mathbb B}^{-1}e^{\mathbb G_0 \tau_\mathrm f}\mathbb M_m(\tau_\mathrm f)
    \mathbb N_{n-m}
    \label{RecurrenceRelation}
\end{align}
with $\mathbb N_0=\mathbb I_{2N}$.
From Eqs.~(\ref{solution}) and (\ref{SolutionBC}), the solution satisfying the boundary conditions is given as follows:
\begin{align}
&\begin{pmatrix}
    \mathbf x_n\\\mathbf v_n\gamma^{-1}
    \end{pmatrix}
    =\lambda^ne^{\mathbb G_0 \tau}\mathbb T_n(\tau)
\hat {\mathbb B}^{-1}
\begin{pmatrix}
    \mathbf x^\mathrm i\\\mathbf x^\mathrm f
    \end{pmatrix},\\
&\mathbb T_n(\tau):=\sum_{m=0}^n\left[\mathbb M_m(\tau)\mathbb N_{n-m}\right].
\end{align}

By substituting the above solutions into the modified Onsager--Machlup integral Eq.~(\ref{ModOMIntegral}) and by comparing with Eq.~(\ref{ResultCGF}), we obtain the following formal expression of $\mathbb E_n$:
\begin{align}
&\mathbb E_n=\hat{\mathbb B}^{-1,\mathrm T}\sum_{m=0}^n\left[\int_0^{\tau_\mathrm f} d\tau\,\mathbb T_m^\mathrm T(\tau)e^{\mathbb G_0^\mathrm T\tau}\mathbb R_0e^{\mathbb G_0\tau}\mathbb T_{n-m}(\tau)\right]\hat{\mathbb B}^{-1}\nonumber\\
&+\hat{\mathbb B}^{-1,\mathrm T}\sum_{m=0}^{n-1}\left[\int_0^{\tau_\mathrm f} d\tau\,\mathbb T_m^\mathrm T(\tau)e^{\mathbb G_0^\mathrm T\tau}\mathbb R_1e^{\mathbb G_0\tau}\mathbb T_{n-m-1}(\tau)\right]\hat{\mathbb B}^{-1}.
\label{GeneralResultE}
\end{align}
This expression can be used to calculate the cumulant generating function for a specific parameter set of $\mathbb L$ and $\mathbb K$. Although the simplest two-component system is discussed in Sec.~\ref{TCS}, concrete calculations for other systems are left for future work.

\section{Two-component system}
\label{TCS}

To demonstrate the influence of odd elasticity on the irreversibility, we consider a two-component system, i.e., $N=2$, and we calculate the cumulant generating function analytically.
We also assume the simplest case, i.e.,
$\mathbb L=\mathbb I_2, \mathbb S=\mathbb I_2$, and \begin{align}
\mathbb A=\begin{pmatrix}
0&1\\
-1&0
\end{pmatrix},
\end{align}
which corresponds mathematically to the charged Brownian particle in the magnetic field~\cite{Aquino10,Jayannavar07,Saha08,Aquino09}.

Then, the extremum equation Eq.~(\ref{pExtremumEq}) is reduced to:
\begin{align}
\ddot {\mathbf x}\gamma^{-2}+2(1+2\xi)\lambda\mathbb A\dot {\mathbf x}\gamma^{-1}-(1+\lambda^2)\mathbf x=0,
\end{align}
which we obtain by modulating an existing equation~\cite{YKLHSK22} with $\xi$.
A solution to this linear equation is given as follows:
\begin{align}
\mathbf x(\tau)=\sum_n^4c_n\mathbf u_ne^{g_n \tau},
\end{align}
where $\mathbf u_n$ and $g_n$ are the corresponding eigenvector and eigenvalue, respectively:
\begin{align}
&\mathbf u_1=(1,i)^\mathrm T,~~~g_1=\sqrt{1-4\lambda^2\xi(1+\xi)}-i\lambda(1+2\xi),\\
&\mathbf u_2=\mathbf u_1,~~~g_2=-g_1^*,\\
&\mathbf u_3=\mathbf u_1^*,~~~g_3=g_1^*,\\
&\mathbf u_4=\mathbf u_1^*,~~~g_4=-g_1,
\end{align}
where star indicates a complex conjugate.
Here, we impose the initial and final conditions for trajectory $\mathbf x(\tau)$ as $\mathbf x^\mathrm i$ and $\mathbf x^\mathrm f$. Then, the solution satisfying these conditions is obtained as follows:
\begin{align}
&\mathbf x(\tau)=-\frac{\sinh(b(\tau-\tau_\mathrm f))}{\sinh(b\tau_\mathrm f)}
\mathbb D^\mathrm T(\tau)\mathbf x^\mathrm i+\frac{\sinh(b\tau)}{\sinh(b\tau_\mathrm f)}
\mathbb D^\mathrm T(\tau-\tau_\mathrm f)\mathbf x^\mathrm f,
\label{sol2d}
\end{align}
where we use the following:
\begin{align}
\mathbb D(\tau):=
\begin{pmatrix}
\cosh(a\tau)&-i\sinh(a\tau)\\
i\sinh(a\tau)&\cosh(a\tau)
\end{pmatrix},
\end{align}
and $a:=i\lambda(1+2\xi), b:=\sqrt{1-4\lambda^2\xi(1+\xi)}$.

\begin{figure}[t]
\begin{center}
\includegraphics[scale=0.5]{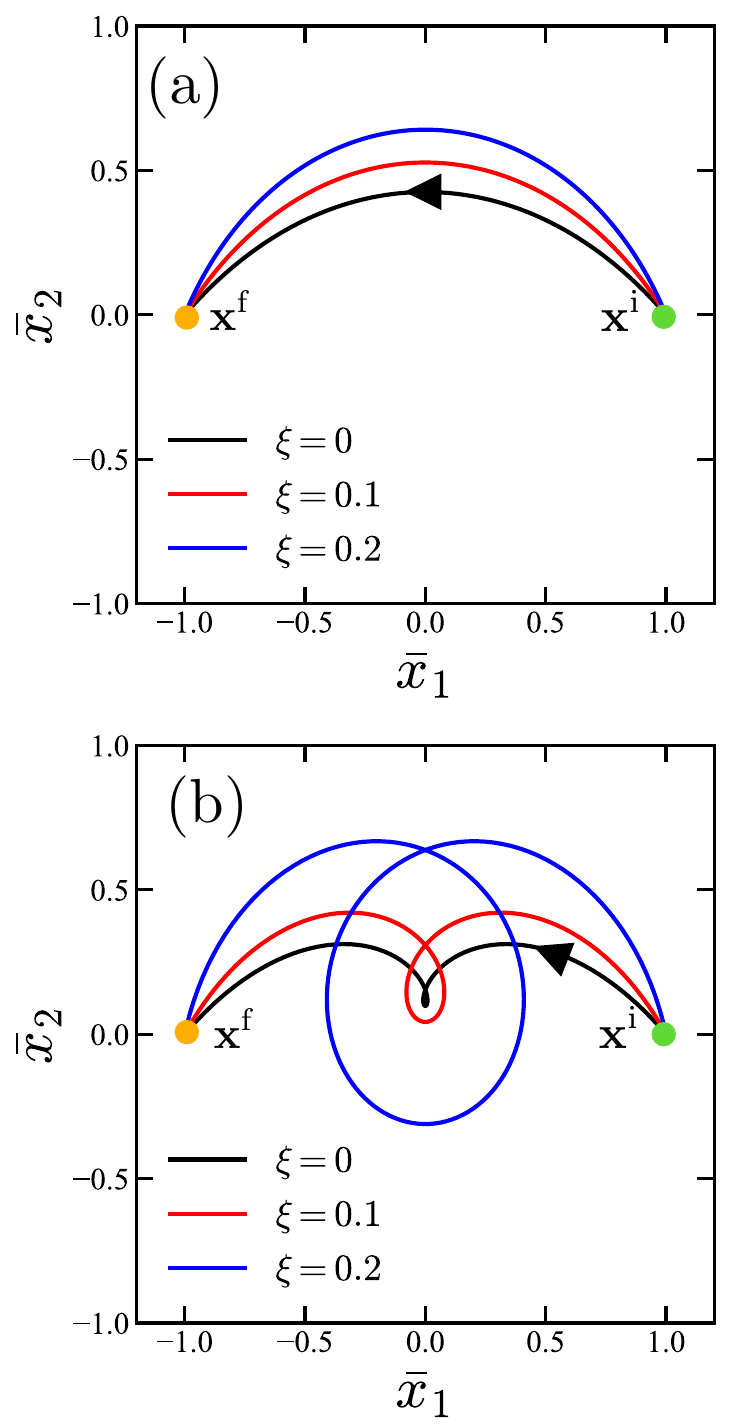}
\end{center}
\caption{
Optimal trajectories (Eq.~(\ref{sol2d})) for $\lambda=1$ and (a) $\tau_\mathrm f=1$, (b) $\tau_\mathrm f=5$. The initial and final states are $\bar{\mathbf x}^\mathrm i=\sqrt{k/k_\mathrm BT}\mathbf x^\mathrm i=(1,0)^\mathrm T$ (green circles) and $\bar{\mathbf x}^\mathrm f=(-1,0)^\mathrm T$ (orange circles), respectively.
Here, the different colors indicate various $\xi$ values.
The variations between the black ($\xi=0$) and other lines implies cumulants.}
\label{Fig:Path}
\end{figure}

In Fig.~\ref{Fig:Path}, we plot the sample trajectories under the conditions $\bar{\mathbf x}^\mathrm i=(1,0)^\mathrm T$ (green circles) and $\bar{\mathbf x}^\mathrm f=(-1,0)^\mathrm T$ (orange circles), where we introduce the dimensionless state variables $\bar{\mathbf x}=\sqrt{k/k_\mathrm BT}\mathbf x$. Here, we use specific values $\lambda=1$ and (a) $\tau_\mathrm f=1$ and (b) $\tau_\mathrm f=5$. 
We vary $\xi$, which are shown using different line colors. 
Eventually, the variations between the black and other lines lead to cumulants.

\begin{figure}[t]
\begin{center}
\includegraphics[scale=0.5]{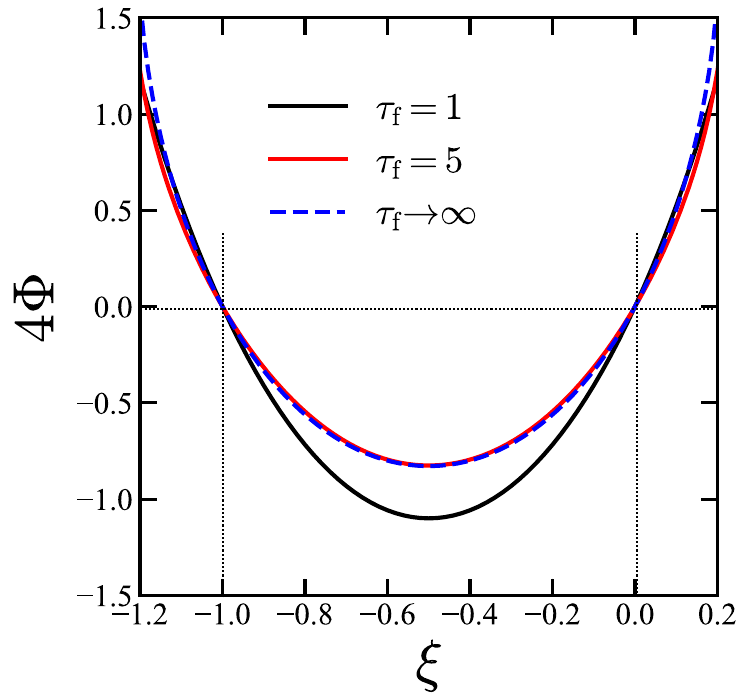}
\end{center}
\caption{
Cumulant generating function (Eq.~(\ref{QGF2d})) for $\lambda=1$. The initial and final states are $\bar{\mathbf x}^\mathrm i=(1,0)^\mathrm T$ and $\bar{\mathbf x}^\mathrm f=(-1,0)^\mathrm T$.
We vary the duration time as $\tau_\mathrm f=1,5$ and $\tau_\mathrm f\to\infty$, which are shown using solid black, solid red, and dashed blue lines, respectively.
Note that the solid red and dashed blue lines nearly coincide.
All lines hold the normalization condition $\Phi(0)=0$.
The derivatives at $\xi=0$ are the corresponding cumulants.
$\Phi(-1)=0$ is shown in all cases and indicates that this transition does not change the system entropy $s_\mathrm{sys}$, as discussed in Sec.~\ref{Dis}.
}
\label{Fig:Q}
\end{figure}

By substituting the solution given by Eq.~(\ref{sol2d}) into Eq.~(\ref{ModOMIntegral}), we calculate the cumulant generating function $\Phi(\xi)$ as follows:
\begin{align}
&\Phi(\xi)=C-\frac{(1+2\xi)}{2k_\mathrm BT}\Delta U-\frac{k}{4k_\mathrm BT}\begin{pmatrix}
    \mathbf x^\mathrm i\\\mathbf x^\mathrm f
    \end{pmatrix}^\mathrm T\mathbb E\begin{pmatrix}\mathbf x^\mathrm i \\ \mathbf x^\mathrm f\end{pmatrix},
\label{QGF2d}
\end{align}
where
\begin{align}
\mathbb E=\frac{b}{\sinh(b\tau_\mathrm f)}\begin{pmatrix}\cosh(b\tau_\mathrm f)\mathbb I_2&
-\mathbb D(\tau_\mathrm f)\\
-\mathbb D^\mathrm T(\tau_\mathrm f)&\cosh(b\tau_\mathrm f)\mathbb I_2
\end{pmatrix},
\label{QGF2dE}
\end{align}
and $C$ is a $\xi$-independent constant determined by an normalization condition $\Phi(0)=0$.
Unlike the formal result given by Eq.~(\ref{ResultCGF}) in Sec.~\ref{Model}, which involves expansion around $\xi=0$, the result given by Eq.~(\ref{QGF2d}) is the full expression for the entire $\xi$. Notice $\mathbb E=\sum_{n=1}\xi^n\lambda^n\mathbb E_n$ for consistency between Eqs.~(\ref{ResultCGF}) and (\ref{QGF2d}). 
In Eq.~(\ref{QGF2d}), $\Phi(\xi;\mathbf x^\mathrm i,\mathbf x^\mathrm f)=\Phi(-1-\xi;\mathbf x^\mathrm f,\mathbf x^\mathrm i)$, which is known as the Gallavotti--Cohen symmetry~\cite{Seifert12} equivalent to the fluctuation theorem.
We obtain each cumulants by expanding Eq.~(\ref{QGF2d}) as Eq.~(\ref{CumulantsQGF}).
Concrete expressions of the first and second cumulants are given in App.~\ref{AppB}.

In Fig.~\ref{Fig:Q}, we plot the cumulant generating function $\Phi(\xi)$, i.e., Eq.~(\ref{QGF2d}), under the conditions $\bar{\mathbf x}^\mathrm i=(1,0)^\mathrm T$ and $\bar{\mathbf x}^\mathrm f=(-1,0)^\mathrm T$ with $\lambda=1$.
Here, the duration time is varied as $\tau_\mathrm f=1,5$ and $\tau_\mathrm f\to\infty$, and we show them as a solid black line, a solid red line, and a blue dashed line, respectively.
The red solid and blue dashed lines show nearly the same value for the entire $\xi$.
The normalization condition is represented as $\Phi(0)=0$, which holds in all cases.
The slope and curvature at $\xi=0$ represent the first and second cumulants, respectively.

Considering zero odd elasticity $\lambda=0$, we can confirm that Eq.~(\ref{QGF2d}) reduces to Eq.~(\ref{Q0}), and we can obtain $\langle\Sigma^1\rangle_\mathrm c=-\Delta U/T$ and $\langle\Sigma^2\rangle_\mathrm c=\langle\Sigma^3\rangle_\mathrm c=\cdots=0$, as discussed in Section~\ref{Model} for arbitrary $N$.

By taking a long time limit $t_\mathrm f\to\infty$, we obtain the following:
\begin{align}
\mathbb E=\sqrt{1-4\lambda^2\xi(1+\xi)}\mathbb I_4
\label{QGF2dElim}
\end{align}
as a limit of Eq.~(\ref{QGF2dE}), and the cumulant generating function becomes similar to that for the entropy production of the steady-state transverse diffusion system~\cite{Buisson23}, which is mathematically the same as the two-component Langevin system discussed in this paper.
Expanding the long time cumulant generating function (refer to Eqs.~(\ref{CumulantsQGF}) and (\ref{QGF2dElim})), we obtain the following cumulants up to the fourth order:
\begin{align}
\langle\Sigma^1\rangle_\mathrm c&=-\frac{\Delta U}{T}+\frac{k}{2T}\lambda^2[(\mathbf x^\mathrm i)^2+(\mathbf x^\mathrm f)^2],
\label{C1}\\
\langle\Sigma^2\rangle_\mathrm c&=\frac{k_\mathrm Bk}{T}\lambda^2(1+\lambda^2)[(\mathbf x^\mathrm i)^2+(\mathbf x^\mathrm f)^2],
\label{C2}\\
\langle\Sigma^3\rangle_\mathrm c&=\frac{6k_\mathrm B^2k}{T}\lambda^4(1+\lambda^2)[(\mathbf x^\mathrm i)^2+(\mathbf x^\mathrm f)^2],\\
\langle\Sigma^4\rangle_\mathrm c&=\frac{12k_\mathrm B^3k}{T}\lambda^4(1+6\lambda^2+5\lambda^4)[(\mathbf x^\mathrm i)^2+(\mathbf x^\mathrm f)^2].
\end{align}
Note that higher cumulants also can be obtained systematically.
All cumulants increase monotonically with the oddness $\lambda$, which implies that the uncertainty of the state transition is strengthened by the oddness.

\section{Applications to an enzyme}
\label{Enzyme}

\begin{figure*}[t]
\begin{center}
\includegraphics[scale=0.4]{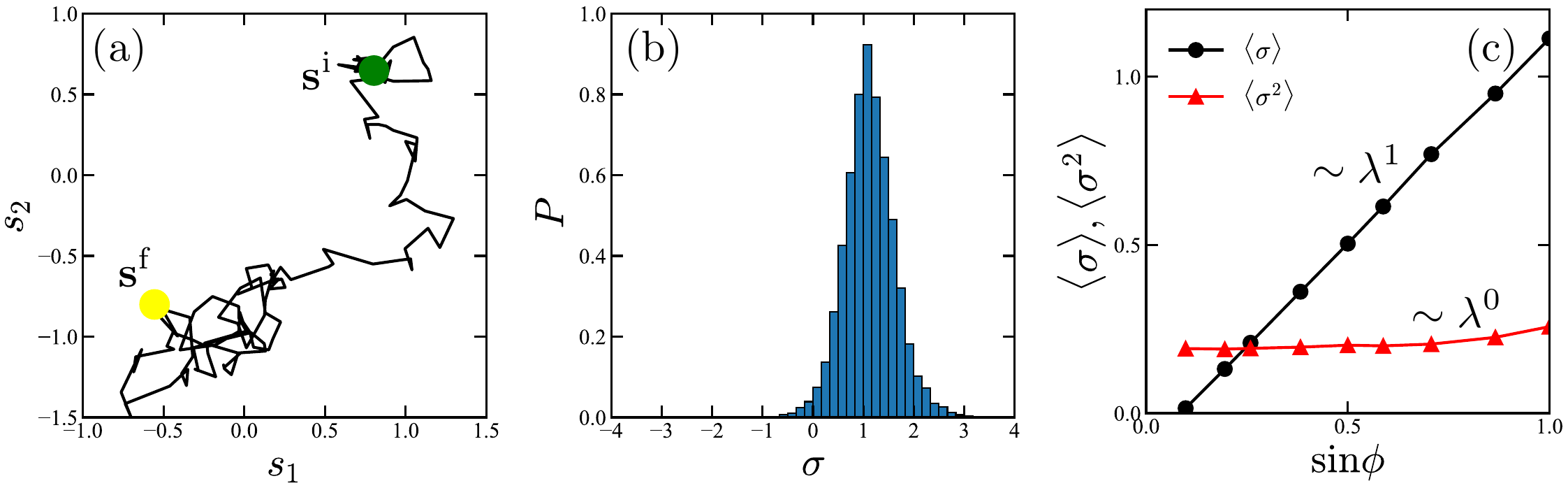}
\end{center}
\caption{
(a) Transition-trajectory from $\mathbf s^\mathrm i$ (green circle) to $\mathbf s^\mathrm f$ (yellow circle) in a state space spanned by $s_1$ and $s_2$. 
(b) Histogram of the scaled-irreversibility $\sigma$ generated by $10000$ transition-trajectories linking $\mathbf s^\mathrm i$ and $\mathbf s^\mathrm f$.
(c) Average and variance of $\sigma$ as functions of $\sin\phi$. It is pointed out that $\sin\phi$ is proportional to the effective odd elasticity in the enzyme model, i.e, $\lambda\sim \sin\phi$~\cite{YIKLSHK22,KYILSHK23}. From this plot, we notice power-law behaviors of the average and the variance as $\langle\sigma\rangle\sim\lambda^1$ and $\langle\sigma^2\rangle\sim\lambda^0$. Regarding the scaling relation between $\Sigma$ and $\sigma$, this results lead to $\langle\Sigma\rangle\sim\lambda^2$ and $\langle\Sigma^2\rangle\sim\lambda^2$ corresponding to our expectation in Eqs.~(\ref{C1}) and (\ref{C2}).
}
\label{Fig:Enzyme}
\end{figure*}

An important application of this research is the conformation change of an enzyme from the open state to closed state binding substrate, which can be considered as the state transition $\mathbf x^\mathrm i\to\mathbf x^\mathrm f$. To highlight the connection between our results and the enzyme, we introduce the mechano-chemical coupling enzyme model~\cite{Yasuda21a,Canalejo21}, which is one of the simplest models describing a structure change of the enzyme molecule (details of this model are provided in App.\ref{AppD} and Ref.~\cite{Yasuda21a}).
In this model, we introduce an extent of chemical reactions of substrates $\theta(t)$, which experiences a tilted periodic potential describing both an energy barrier and a chemical potential difference between a substrate and a product.
In addition, the enzyme structure is represented by two variables $s_1(t), s_2(t)$ and coupled with $\theta$ through the mechano-chemical coupling energy given in Eq.~(\ref{Gs}).
Note that the mechano-chemical coupling includes a phase difference $\phi$ between $s_1$ and $s_2$.
Then, we employ the Onsager's phenomenological equations with thermal fluctuations (see Eqs.~(\ref{OPE1})-(\ref{OPE3})) to describe dynamics of $\theta(t)$, $s_1(t)$, and $s_2(t)$.
This model has two stable fixed points $\mathbf s^\mathrm i$ and $\mathbf s^\mathrm f$, representing open and closed states, in $s_1$-$s_2$ space.

Solving the model equations numerically with specific parameters written in App.\ref{AppD}, 
we observe state transition between two stable fixed points $\mathbf s^\mathrm i\to\mathbf s^\mathrm f$ with $\theta$ increasing by $\pi$.
One of the transition-trajectories is shown in Fig.~\ref{Fig:Enzyme}(a).
Then, we collect $10000$ transition-trajectories regardless of the duration time $t_\mathrm f$ and build an ensemble of the transition-trajectories.
In order to connect the ensemble and our results on $\Sigma$, we introduce a scaled-irreversibility,
\begin{align}
\sigma=\int_0^\tau dt(\dot s_1s_2-\dot s_2s_1),
\end{align}
which can be directly calculated from the numerical trajectories and has a scaling relation $\sigma\sim\Sigma/(k\lambda)$.
Then, we generate a histogram of the scaled-irreversibility $\sigma$ using the obtained ensemble in Fig.~\ref{Fig:Enzyme}(b).
From this histogram, we calculate average and variance of the scaled-irreversibility and plot them as functions of $\sin\phi$ in Fig.~\ref{Fig:Enzyme}(c).
It is reported in Ref.~\cite{YIKLSHK22,KYILSHK23} that estimated oddness $\lambda$ in this enzyme model is proportional to the sine of the phase difference, i.e., $\lambda\sim\sin\phi$.
Basing on this scaling, we can show $\langle\sigma\rangle\sim\lambda^1$ and $\langle\sigma^2\rangle\sim\lambda^0$ from Fig~\ref{Fig:Enzyme}(c).
Finally, recalling the scaling relation $\sigma\sim\Sigma/(k\lambda)$, we get $\langle\Sigma\rangle\sim\lambda^2$ and $\langle\Sigma^2\rangle\sim\lambda^2$ which qualitatively correspond to the small $\lambda$ limit of our expectation in Eqs.~(\ref{C1}) and (\ref{C2}).

This discussion shows that the linear Langevin equation Eq.~(\ref{OddLangevinSystem}) can be applied to enzymes for the qualitative scaling relations.
Nevertheless, nonlinearity is also needed to reproduce the complexity of the enzyme dynamics. Attempts to introduce nonlinearity to  a system with odd elasticity have been reported in microswimmers~\cite{Ishimoto22,Yasuda21b,Ishimoto23}.
Based on these studies, the role of nonlinearity in an enzyme system can be investigated more extensively.

In addition to the above qualitative discussion, we provide rough estimation of parameters.
The energy injected into the enzyme is part of the chemical potential difference between the substrate and the products, e.g., the chemical potential difference between the ATP and ADP molecule is roughly $20k_\mathrm BT$~\cite{Toyabe10}.
Here, we assume that the enzyme receives 10\% of the chemical potential difference, i.e., $2k_\mathrm BT$, during the reaction.
In the Langevin system with the odd elasticity, the injected energy is represented by $k\lambda a^2$, where $a$ is the size of the object, e.g., $10\mathrm{nm}$ for enzyme molecule.
Thus, the odd elasticity can be estimated roughly as $k\lambda\sim 2k_\mathrm BT/(10\mathrm{nm})^2\sim10^{-4}\mathrm {J/m^2}$.
The duration time for protein folding is roughly estimated as $t_\mathrm f\sim 1\mathrm{\mu s}$~\cite{Chung12}.

\section{Summary and Discussion}
\label{Dis}

In this paper, we have calculated the cumulant generating function of the irreversibility of the state transition $\mathbf x^\mathrm i\to\mathbf x^\mathrm f$ in the Langevin system with odd elasticity using the path probability determined by the Onsager--Machlup integral and the framework of the large deviation theory.
As a result, for the $N$-component system, we have provided a formal expression of the cumulant generating function in Eq.~(\ref{ResultCGF}), which indicates that the oddness leads to a higher-order cumulant unlike a passive elastic system, where the irreversibility is determined by the energy change of the system without error.

To demonstrate the effect of oddness using the concrete parameter setup, we further calculated the cumulant generating function of a simple two-component system. Here, we found that the cumulants increases monotonically with oddness $\lambda$, which implies that the oddness amplifies the uncertainty of the system.
The oddness may help us understand the uncertainty of various biological systems, e.g., random responses to stimulations.

In this study, we have used the approximation expressed in Eq.~(\ref{VaradhanT}) with the assumption that $\hat O\gg k_\mathrm BT$~\cite{Touchette09}. Because it can be supposed that $\hat O$ increases with the system size $N$ and duration time $t_\mathrm f$, either $N$ or $t_\mathrm f$ is implicitly assumed to be large.
These assumptions are valid in macroscopic systems with many degrees of freedom but unclear for relatively small systems. Hence, we should be careful on the assumption that $\hat O\gg k_\mathrm BT$ for small systems. 

Kwon \textit{et al.}\ provide the method to calculate the cumulant generating function without the above assumption for an unfixed initial and a fixed final conditions~\cite{Kwon11}. They calculate the path integral by performing the Gaussian integrals for each discretized time step in a recurrent manner and obtain a time evolution equation of a kernel $\mathbb E^{\mathrm f\mathrm f}(t)$, which contributes the cumulant generating function with a form of $\mathbf x^\mathrm f \mathbb E^{\mathrm f\mathrm f}(t_\mathrm f)\mathbf x^\mathrm f$. We can consider applying this method to the state transitions between fixed initial and final state and improve the precision of the current work. 
In the case of the state transition, a calculation of the kernel $\mathbb E^{\mathrm i\mathrm f}(t)$, which links the initial and final stats and contributes the cumulant generating function with the form of $\mathbf x^\mathrm i \mathbb E^{\mathrm i\mathrm f}(t_\mathrm f)\mathbf x^\mathrm f$ (see Eq.~(\ref{QGF2d})), is a central problem. However, obtaining $\mathbb E^{\mathrm i\mathrm f}$ is expected to be more difficult than $\mathbb E^{\mathrm f\mathrm f}$ because the state transition is a two-time boundary problem.
Moreover, their recurrent calculations will be more complicated because $\mathbb E^{\mathrm i\mathrm f}$ involves knowledge of all time steps, including the initial state, unlike $\mathbb E^{\mathrm f\mathrm f}$.  
Hence, calculations extended beyond the saddle point approximation are not straightforward but should be performed in the future.

In the following, we discuss the connection between our results and stochastic thermodynamics.
The correspondence between the irreversibility $\Sigma$ and the thermodynamic entropy of a thermal bath (or heat from a system to  a bath $Q$)~\cite{Seifert05} is not clarified in active systems with odd elasticity because there are two choices of reversed trajectories in which oddness is flipped or not~\cite{Fodor22,Byrne22}.
This issue is similar to the time-reversal symmetry of a system in a magnetic field.
However, unlike the magnetic field, the microscopic origin of the odd elasticity is not identified, which complicates the problem.
In addition, the approximation $\hat O\gg k_\mathrm BT$ used in this paper also obscures the connection to thermodynamics.
In addition to the irreversibility $\Sigma$, we consider the system entropy $s_\mathrm {sys}(\mathbf x)=-k_\mathrm B\ln \mathcal P(\mathbf x)$, where $\mathcal P(\mathbf x)$ is a steady-state distribution.
In the Langevin system given by Eq.~(\ref{OddLangevinSystem}), $\mathcal P(\mathbf x)$ is the Gaussian distribution $\mathcal P(\mathbf x)\sim\exp[-\mathbf x^\mathrm T\mathbb C^{-1}\mathbf x/2]$, where $\mathbb C$ is a covariance matrix obtained by solving the Lyapunov equation $k\mathbb L\mathbb K\mathbb C+k\mathbb C\mathbb K^\mathrm T\mathbb L=2k_\mathrm BT \mathbb L$~\cite{YIKLSHK22}.
Note that the cumulant generating function $\Phi(\xi)$ and system entropy $s_\mathrm {sys}$ are connected as $\Phi(-1)=\ln\langle e^{-\Sigma/k_\mathrm B}\rangle_{\mathrm i\to \mathrm f}=\Delta s_\mathrm {sys}/k_\mathrm B$ under the assumption of $\Sigma=Q/T$, which is an alternative form of the Jarzynski equality~\cite{Seifert12} (the derivation of this equality is given in App.~\ref{AppC}).
Thus, the cumulant generating function has information on both the irreversibility and the system entropy.
In Fig.~\ref{Fig:Q}, we observe $\Phi(-1)=0$, indicating that the system entropy does not change, i.e., $\Delta s_\mathrm {sys}=0$, during the $(1,0)\to(-1,0)$ transition, which is symmetric between the initial and final states. Notably, $\Delta s_\mathrm {sys}\ne0$ for other transitions in general.

In Section~\ref{Enzyme}, we have discussed enzymes as a physical system, as represented by the Langevin system with odd elasticity.
In addition to this example, this model can be used to describe other nonequilibrium physical systems, e.g., charged Brownian particle in a magnetic field~\cite{Aquino10,Jayannavar07,Saha08,Aquino09}, the polymer molecule in an external shear flow~\cite{Turitsyn07}, and the transverse diffusion system~\cite{Noh13,Noh14,Buisson23}.
A previous study~\cite{Buisson23} reported the cumulant generating function and rate function of the steady-state entropy production of the transverse diffusion system, which is mathematically the same as two-component Langevin system with odd elasticity (Section\ref{TCS}).
The obtained cumulant generating function is $\Phi(\xi)=1-\sqrt{1-4\xi(1+\xi)\lambda^2}$~\cite{Buisson23}.
Although this is similar to our results for the long time limit given by Eq.~(\ref{QGF2dElim}), differing from our results, their results for the steady-state do not include the initial and final states.
Note that we consider transitions rather than the steady state considered in the literature~\cite{Buisson23}.

In this study, we calculated the cumulant generating function of irreversibility using the Onsager--Machlup integral.
The method of the Onsager--Machlup integral and the large deviation theory have diverse applications.
One such application is the calculation of the conditioned distribution $\mathcal P(\mathbf x, t|\mathbf x^\mathrm i)$ through the optimal problem of the Onsager--Machlup integral~\cite{Majumdar20}.
Another application is the estimation of the mean passage time, which is an average value of the duration time $\langle t_\mathrm f\rangle$~\cite{Woillez19}. 
These quantities will be calculated in the Langevin system with odd elasticity.

K.Y.\ thanks K.\ Ishimoto and S.\ Komura for meaningful discussions.
K.Y.\ acknowledges support by a Grant-in-Aid for JSPS Fellows (Grants No.\ 22KJ1640) from the JSPS
and the Research Institute for Mathematical Sciences, an International Joint Usage/Research Center located in Kyoto University.
K.Y.\ would like to thank Enago (www.enago.jp) for the English language review.

\appendix
\section{Derivation of recurrence relation Eq.~(\ref{RecurrenceRelation})}
\label{AppA}
Here, we present the derivation of the recurrence relation given in Eq.~(\ref{RecurrenceRelation}).
Using Eq.~(\ref{solution}), the boundary conditions $\mathbf x_n(0)=\mathbf x_n(t_\mathrm f)=0$ can be rewritten as follows:
\begin{align}
\begin{pmatrix}
    0\\0\\-\sum_{m=1}^n\left[\lambda^me^{\mathbb G_0 \tau_\mathrm f}\mathbb M_m(\tau_\mathrm f)
    \begin{pmatrix}
    \mathbf c_{n-m}\\\mathbf d_{n-m}
    \end{pmatrix}\right]
    \end{pmatrix}
    =\mathbb B
        \begin{pmatrix}
    \mathbf c_n\\\mathbf d_n\\\mathbf v^\mathrm i\gamma^{-1}\\\mathbf v^\mathrm f\gamma^{-1}
    \end{pmatrix}.
\end{align}
   By solving this equation, we obtain:
   \begin{align}
    \begin{pmatrix}
    \mathbf c_n\\\mathbf d_n
    \end{pmatrix}=-
    \tilde {\mathbb B}^{-1}
\sum_{m=1}^n\left[\lambda^me^{\mathbb G_0 \tau_\mathrm f}\mathbb M_m(\tau_\mathrm f)
    \begin{pmatrix}
    \mathbf c_{n-m}\\\mathbf d_{n-m}
    \end{pmatrix}\right].
\end{align}
By substituting Eq.~(\ref{SolutionBC}) into the above equation, we obtain the recurrence relation Eq.~(\ref{RecurrenceRelation}).

\section{First and second cumulants for two-component system}
\label{AppB}
In the following, we present the explicit form of the first and second cumulants for the two-component system.
Recalling Eq.~(\ref{CumulantsQGF}), the first and second cumulants can be obtained via expansion of the cumulant generating function given in Eq.~(\ref{QGF2d}).

Here, the first cumulant is given as follows:
\begin{align}
\langle\Sigma^1\rangle_\mathrm c&=-\frac{\Delta U}{T}-\frac{k\lambda}{4T}\begin{pmatrix}
    \mathbf x^\mathrm i\\\mathbf x^\mathrm f
    \end{pmatrix}^\mathrm T \begin{pmatrix}\mathbb E_1^\mathrm d&\mathbb E_1^\mathrm {od} \\(\mathbb E_1^\mathrm {od})^\mathrm T&\mathbb E_1^\mathrm d\end{pmatrix}\begin{pmatrix}\mathbf x^\mathrm i \\ \mathbf x^\mathrm f\end{pmatrix},
\end{align}
where we use the following matrixes:
\begin{align}
\mathbb E_1^\mathrm d=\frac{2}{\sinh^2(\tau_\mathrm f)}\lambda(\tau_\mathrm f-\cosh(\tau_\mathrm f)\sinh(\tau_\mathrm f))\mathbb I_2,
\end{align}
\begin{align}
&\mathbb E_1^\mathrm {od}=\frac{2}{\sinh^2(\tau_\mathrm f)}
[-\lambda(\tau_\mathrm f\cosh(\tau_\mathrm f)-\sinh(\tau_\mathrm f))\mathbb D_0\nonumber\\
&+\tau_\mathrm f\sinh(\tau_\mathrm f)\hat {\mathbb D}_0].
\end{align}

In the same manner, we obtain the second cumulant as follows:
\begin{align}
&\langle\Sigma^2\rangle_\mathrm c=-\frac{2k_\mathrm Bk\lambda^2}{4T}\begin{pmatrix}
    \mathbf x^\mathrm i\\\mathbf x^\mathrm f
    \end{pmatrix}^\mathrm T\begin{pmatrix}\mathbb E_2^\mathrm d&\mathbb E_2^\mathrm {od} \\(\mathbb E_2^\mathrm {od})^\mathrm T&\mathbb E_2^\mathrm d\end{pmatrix}\begin{pmatrix}\mathbf x^\mathrm i \\ \mathbf x^\mathrm f\end{pmatrix},
\end{align}
where
\begin{align}
&\mathbb E_2^\mathrm d=\frac{2\mathbb I_2}{\sinh^2(\tau_\mathrm f)}[\tau_\mathrm f(1-\lambda^2)\nonumber\\
&+\cosh(\tau_\mathrm f)\sinh(\tau_\mathrm f)(2\lambda^2\tau_\mathrm f^2/\sinh^2(\tau_\mathrm f)-1-\lambda^2)],
\end{align}
\begin{align}
&\mathbb E_2^\mathrm {od}=\frac{2}{\sinh^3(\tau_\mathrm f)}[-[-(1+\lambda^2)\sinh^2(\tau_\mathrm f)\nonumber\\
&+(1-\lambda^2)\tau_\mathrm f\cosh(\tau_\mathrm f)\sinh(\tau_\mathrm f)\nonumber\\
&+\lambda^2\tau_\mathrm f^2(\cosh^2(\tau_\mathrm f)+1)-\tau_\mathrm f^2\sinh^2(\tau_\mathrm f)]\mathbb D_0 \nonumber\\
&+2\lambda\tau_\mathrm f[\tau_\mathrm f\cosh(\tau_\mathrm f)-\sinh(\tau_\mathrm f)]\hat{\mathbb D}_0].
\end{align}
In addition, we use the following: 
\begin{align}
&\mathbb D_0:=
\begin{pmatrix}
\cos(\lambda \tau_\mathrm f)&\sin(\lambda \tau_\mathrm f)\\
-\sin(\lambda \tau_\mathrm f)&\cos(\lambda \tau_\mathrm f)
\end{pmatrix},\\
&\hat{\mathbb D}_0:=
\begin{pmatrix}
\sin(\lambda \tau_\mathrm f)&-\cos(\lambda \tau_\mathrm f)\\
\cos(\lambda \tau_\mathrm f)&\sin(\lambda \tau_\mathrm f)
\end{pmatrix}.
\end{align}
Note that we can also obtain higher cumulants systematically.

\section{Mechano-chemical coupling enzyme model}
\label{AppD}

In this appendix, we show details of the enzyme model that was proposed considering mechano-chemical coupling in Refs.~\cite{Yasuda21a,Canalejo21}.

In this model, we consider the extent of the catalytic reaction 
$\theta(t)$ and the structure of an enzyme denoted by $s_1(t)$ and $s_2(t)$.
We introduce the free energy describing a chemical reaction $g_{\mathrm{r}}(\theta)=-h\cos(2\theta)-\nu\theta$, where $h$ is the energy barrier and $\nu$ is the chemical potential difference.
We also use the mechano-chemical coupling energy:
\begin{align}
&g_\mathrm{c}(\theta,\{s_i\})=\frac{c}{2}\left([s_1-\sin(\theta+(p-\phi)/2)]^2\right.\nonumber\\
&\left.+[s_2-\sin(\theta+(p+\phi)/2)]^2\right).
\label{Gs}
\end{align}
Here, $c$ is the coupling strength, $d$ is the amplitude of the structure change, $p$ is the phase difference relative to the reaction phase, and $\phi$ is the phase difference between $s_1$ and $s_2$.
In this paper, we use the fixed phase difference, $p=\cos^{-1}(\nu/h)$.
Then, the total free energy is $g_{\mathrm{t}}(\theta,\{s_i\})=g_\mathrm{r}(\theta)+g_\mathrm{c}(\theta,\{s_i\})$.
We employ the Onsager's phenomenological equations for dynamics of $\theta(t)$, $s_1(t)$, and $s_2(t)$:
\begin{align}
\dot{\theta}& =-\mu_{\theta}\partial_\theta g_{\mathrm{t}}+\sqrt{2\mu_\theta}\xi, \label{OPE1}
\\
\dot{s_1}& =-\mu_s\partial_{s_1} g_{\mathrm{t}}+ \sqrt{2\mu_s}\xi_1\label{OPE2}\\
\dot{s_2}& =-\mu_s\partial_{s_2} g_{\mathrm{t}}+ \sqrt{2\mu_s}\xi_2\label{OPE3}
\end{align}
Here, $\xi$, $\xi_1$, and $\xi_2$ represent thermal fluctuations that satisfy the fluctuation dissipation theorem~\cite{KuboBook,DoiBook}, i.e., $\langle \xi(t)\xi(0)\rangle=2k_\mathrm BT\mu_{\theta}\delta (t)$ and $\langle \xi_1(t)\xi_1(0)\rangle=\langle \xi_2(t)\xi_2(0)\rangle=2k_\mathrm BT\mu_s\delta (t)$. 
A more detailed explanation of the enzyme model is provided in Ref.~\cite{Yasuda21a}.

Stable fixed points $\mathbf s^\mathrm i$ and $\mathbf s^\mathrm f$ of this model are given by $\mathbf s^\mathrm i=(\sin(\theta^\mathrm i+(p-\phi)/2),\sin(\theta^\mathrm i+(p+\phi)/2))$ and $\mathbf s^\mathrm f=-\mathbf s^\mathrm i$ where $\theta^\mathrm i=\sin^{-1}(\nu/h)$. Note that the transition between the fixed points occurs under activations of the thermal noise.

In numerical calculations shown in Fig~\ref{Fig:Enzyme}, we used the following parameter values: $\nu/(k_\mathrm BT)=8$, $h/(k_\mathrm BT) = 10$, $c/(k_\mathrm BT) = 10$, $\mu_s/\mu_\theta =1$, and $\phi=\pi/2$.

\section{Jarzynski equality}
\label{AppC}
Here, we review the Jarzynski equality discussed in Section~\ref{Dis}.
We consider the first law of thermodynamics and assume the correspondence between the irreversibility and the heat, i.e., $W=\Delta U+Q=\Delta U+T\Sigma$, where $W$ is work done by the system, and $Q$ is heat from the system to the bath.
By introducing the free energy $\Delta F=\Delta U-T\Delta s_\mathrm {sys}$, the Jarzynski equality can be formed as $\langle e^{- W/(k_\mathrm BT)}\rangle_{\mathrm i\to \mathrm f}=e^{-\Delta F/(k_\mathrm BT)}$.
This can be written as $\langle e^{- \Sigma/k_\mathrm B}\rangle_{\mathrm i\to \mathrm f}=e^{-\Delta s_\mathrm {sys}/k_\mathrm B}$, which leads to the second law
 $\Delta s_\mathrm {sys}+\langle\Sigma\rangle_{\mathrm i\to \mathrm f} \geq 0$ through Jensen's inequality.
   
In the following, we show proof of the Jarzynski equality.
According to the definition of the path averaging, we obtain:
 \begin{align}
\langle e^{-\Sigma/k_\mathrm B} \rangle_{\mathrm i\to \mathrm f}&=\int_{\mathbf x^\mathrm i}^{\mathbf x^\mathrm f} \mathcal D\mathbf x\,e^{-\Sigma/k_\mathrm B} P[\mathbf x(t)|\mathbf x^\mathrm i,\mathbf x^\mathrm f].
\end{align}
Using Bayes' theorem, we obtain the following:
 \begin{align}
\langle e^{-\Sigma/k_\mathrm B} \rangle_{\mathrm i\to \mathrm f}
&=\mathcal P(\mathbf x^\mathrm f|\mathbf x^\mathrm i)^{-1}\int_{\mathbf x^\mathrm i}^{\mathbf x^\mathrm f} \mathcal D\mathbf x\,e^{-\Sigma/k_\mathrm B} P[\mathbf x(t)|\mathbf x^\mathrm i].
\end{align}
Then, we use the definition of $\Sigma$,
 \begin{align}
\langle e^{-\Sigma/k_\mathrm B} \rangle_{\mathrm i\to \mathrm f}
&=\mathcal P(\mathbf x^\mathrm f|\mathbf x^\mathrm i)^{-1}\int_{\mathbf x^\mathrm i}^{\mathbf x^\mathrm f} \mathcal D\mathbf x\,P[\mathbf x^{\mathrm {rev}}(t)|\mathbf x^\mathrm f].
\end{align}
Here, the path integral on the right-hand side becomes a conditioned probability as follows:
 \begin{align}
\langle e^{-\Sigma/k_\mathrm B} \rangle_{\mathrm i\to \mathrm f}
&=\mathcal P(\mathbf x^\mathrm f|\mathbf x^\mathrm i)^{-1}\mathcal P(\mathbf x^\mathrm i|\mathbf x^\mathrm f)
\end{align}
Finally, we utilize the property of conditioned probabilities to obtain the following:
 \begin{align}
\langle e^{-\Sigma/k_\mathrm B} \rangle_{\mathrm i\to \mathrm f}
&=\mathcal P(\mathbf x^\mathrm f)^{-1}\mathcal P(\mathbf x^\mathrm i)=e^{\Delta s_\mathrm {sys}/k_\mathrm B},
\end{align}
which indicates the direct connection between the cumulant generating function of $\Sigma$ and the system entropy change $\Delta s_\mathrm {sys}$.


\end{document}